\documentclass[preprint]{ptephy_v1}

\usepackage{amsmath}
\usepackage{amsthm} 
\usepackage{amssymb}
\usepackage{bm}
\usepackage{graphics}
\usepackage{braket}

\begin{document}

\title{Laplace expansion method for the calculation of the reduced width amplitudes}

\author{Yohei Chiba}
\affil{Department of Physics, Hokkaido University, 060-0810 Sapporo, Japan
\email{chiba@nucl.sci.hokudai.ac.jp}}
\author[1,2]{Masaaki Kimura}
\affil{Nuclear Reaction Data Centre,  Hokkaido University,
Sapporo 060-0810, Japan} 

\begin{abstract}%
We derive the equations to calculate the reduced width amplitudes (RWA) of the different size
clusters and deformed clusters without any approximation. These equations named Laplace expansion 
method are applicable to the nuclear models which uses the Gaussian wave packets. The
advantage of the method is demonstrated by the numerical calculations of the 
${}^{16}{\rm O}+\alpha$ and ${}^{24}{\rm Mg}+\alpha$  RWAs  in $^{20}{\rm Ne}$ and 
$^{28}{\rm  Si}$.  
\end{abstract}

\subjectindex{xxxx, xxx}

\maketitle

\section{Introduction}
In the excited states of light stable nuclei, it is well known that various cluster states appear
as illustrated in Ikeda diagram \cite{Ikeda1968}. They are composed of $\alpha$, $^{12}{\rm C}$
and $^{16}{\rm O}$ clusters which are tightly bound and stable compared to the neighboring
nuclei. In these decades, the study of nuclear clustering is extended to unstable nuclei where
novel types of clustering were found. The molecular-orbit and atomic-orbit states in Be isotopes  
\cite{Seya1981,VonOertzen1996,Kanada-Enyo1999,Itagaki2000,Descouvemont2002,Freer1999,Freer2001,Curtis2004,Milin2005,Bohlen2007}
are the representative of such novel types of clustering. In contrast to the clustering of light
stable nuclei, they are composed of $^{6}{\rm He}$ and $^{8}{\rm He}$ which are weakly bound and
unstable.  Since the definition of the cluster is extended from the ordinary one, we need a good
measure for these non-conventional clustering. 

The reduced width amplitude is one of such measures for the clustering.
It is the cluster formation probability at a given inter-cluster distance, and hence, it is
regarded as a direct evidence of the clustering. By the $R$-matrix theory \cite{Descouvemont2010},
the RWA is derived from the width of the cluster states which are experimentally determined by the 
measurement of the cluster decay lifetime, the resonant scattering and the transfer
reactions. Therefore, numerous experiments have been conducted to determine the RWA and to
identify various clusters.  A variety of cluster states in light $p$-$sd$-shell nuclei illustrated
in the Ikeda diagram have been identified from their large decay widths and RWAs
\cite{Nemoto1972,Sunkel1972,Matsuse1973,Fujiwara1980,Descouvemont1987}. 
Several clusters in heavier $pf$-nuclei were established in 1990's when the measurements of the
RWAs played an essential role to identify the $\alpha$ cluster states of $^{40}{\rm Ca}$ and
$^{44}{\rm Ti}$
\cite{Strohbusch1974,Friedrich1975,Betts1977,Fortune1979,Ohkubo1988,Wada1988,Yamaya1993,Yamaya1994,Michel1998,Yamaya1998,Kimura2006}.  
More recently, the $\alpha$ decay property has 
been used in combination with the isoscalar monopole and dipole transitions to identify the
gas-like $\alpha$ cluster states
\cite{Tohsaki2001,Kawabata2007,Kanada-Enyo2007,Wakasa2007,KAWABATA2008,Funaki2008,Itoh2011} and various clusters in $sd$-shell nuclei \cite{Yamada2008,Kawabata2013,Itoh2013,Chiba2015,Kanada-Enyo2016a,Chiba2016}.
The importance of the RWA for the study of exotic clustering in neutron-rich nuclei must  
also be  emphasized. It was an important observable to identify the molecular clustering in Be 
isotopes \cite{Freer1999,Freer2001,Curtis2004,Milin2005}. More recently, the cluster states in
$^{18}{\rm O}$ and $^{22}{\rm Ne}$
\cite{Scholz1972,Descouvemont1985,Descouvemont1988,Rogachev2001,Curtis2002,Goldberg2004,Yildiz2006,Kimura2007,Furutachi2008,Fu2008,Johnson2009,VonOertzen2010}
 and linear-chain states in $^{14,16}{\rm C}$ 
\cite{Soic2003,VonOertzen2004,Price2007,Haigh2008,Suhara2010a,Freer2014,Baba2014,Tian2016,DellAquila2016,Fritsch2016,Baba2016,Yamaguchi2017,Li2017,Baba2017}
are discussed from their $\alpha$ decays and RWAs. Thus, the comparison of the measured RWA with
the theoretical ones is indispensable to establish the cluster formation.  

However, in the theoretical studies, the calculation of RWA for general cluster systems is not
easy due to the antisymmetrization of nucleons belonging to the different clusters. To simplify
the treatment of the antisymmetrization, the ordinary methods for RWA calculation 
\cite{Balashov1959,Honda1965,Horiuchi1972,Horiuchi1977,Kanada-Enyo2003a,Kimura2004a}
often approximate the cluster wave functions with the $SU(3)$ shell model wave functions
\cite{Elliott1958,Elliott1958a} having the common oscillator parameters (the same size of
clusters). Unfortunately, this approximation limits the applicability of the methods. They are
inaccurate when applied to the unequal size clusters  and the clusters which cannot be
approximated by a single $SU(3)$ shell model wave 
function. Typical examples of this may be the $^{16}{\rm O}+\alpha$ and $^{6}{\rm
He}+\alpha$ clusters, {\it i.e.} the size of these clusters are different and a halo nucleus 
$^{6}{\rm He}$ cannot be described by a single $SU(3)$ shell model wave function. Furthermore,
to calculate the RWA of deformed clusters, the ordinary methods demand much computational time
because of the multiple angular momentum projections. Although an approximate method  proposed by
Kanada-En'yo {\it et al.} \cite{Kanada-Enyo2014b} reduced the  computational cost to some extent,
the development of an alternative method for RWA calculation is highly desirable and in need. 

For this purpose, we present a new method for the RWA calculation. We derive the equations which
can calculate the RWA of the different size clusters and deformed clusters without any
approximation. These equations named Laplace expansion  method are applicable to nuclear
models which uses the Gaussian wave packets such as antisymmetrized molecular dynamics (AMD) 
\cite{Kanada-Enyo2003,Kanada-Enyo2012,Kimura2016}.  This paper is organized as follows. In the
next section, we derive the equations of the Laplace expansion method. We also discuss the
advantages and disadvantages of the method compared to the ordinary method. In the section
\ref{sec:result}, we show the numerical results of the  ${}^{16}{\rm O}+\alpha$ and ${}^{24}{\rm
Mg}+\alpha$  RWAs as the examples of the unequal size clusters and deformed clusters. In the final
section, we summarize the present work.  

\section{Laplace expansion method for RWA calculation}\label{sec:formalism}
In this section, we outline a new method to calculate the RWA which utilizes the Laplace expansion
of the matrix determinant. We first introduce the AMD wave function. Then, by using the Laplace
expansion, we show that the AMD wave function of $A$-body system can be decomposed into those of
the subsystems with masses $C_1$ and $C_2$. With this expansion, we derive the equations to
calculate the RWA which we call Laplace expansion method. We also compare the Laplace expansion
method with ordinary one to discuss its advantages and disadvantages. 

\subsection{Wave function of antisymmetrized molecular dynamics}
The wave function of AMD for $A$-body system is a Slater determinant of the Gaussian wave packets
describing nucleons. 
\begin{align}
 \Psi_{A}^{AMD} &= \frac{1}{\sqrt{C_1!}}  \left| 
 \begin{matrix}
  \braket{\bm{r}_1|\psi_1} &  \dots & \braket{\bm{r}_1|\psi_A} \\
  \vdots &  \ddots & \vdots\\
  \braket{\bm{r}_A|\psi_1} & \dots & \braket{\bm{r}_{A}|\psi_{A}}
 \end{matrix}\right|,\label{eq:amdwf1}\\
 \braket{\bm{r}|\psi_i} &= \left(\frac{|2M|}{\pi^3}\right)^{1/4}
 \exp \left\{ -(\bm{r}-\bm{Z}_i)^T M (\bm{r}-\bm{Z}_i) \right\} 
 (\alpha_i \chi_\uparrow + \beta_i \chi_\downarrow) \eta_i,
\end{align}
where the Gaussian centroids $\bm{Z}_i$ are the complex valued three dimensional vector, and  the
spin directions are parameterized by the complex variables $\alpha_i$ and $\beta_i$. The isospin
part $\eta_i$ is fixed to either of proton or neutron. Each nucleon wave packet has these
independent variables. To discuss the general case, we assume the use of the deformed Gaussian
wave packets \cite{Kimura2004a}, and hence, $M$ denote a symmetric positive-definite $3 \times 3$
matrix.   

If the width matrix $M$ is common for all nucleon wave packets, the AMD wave function can be
straightforwardly decomposed into the internal wave function $\Psi_A^{int}$ and the center-of-mass
wave function $\Psi_{A}^{cm}$, 
\begin{align}
 \Psi_A^{AMD} &= \Psi_A^{int} \Psi_{A}^{cm},\label{eq:cm1}\\
 \Psi_{A}^{cm} &= \left(\frac{|2AM|}{\pi^3}\right)^{1/4}
 \exp\left\{-A(\bm{R}-\bm{Z})^TM(\bm{R}-\bm{Z})\right\},\\
 \bm R &= \frac{1}{A}\sum_{i=1}^A\bm r_i, \quad
 \bm Z = \frac{1}{A}\sum_{i=1}^A\bm Z_i.\label{eq:rc0}
\end{align}
This simple but important decomposition is repeatedly used in the Laplace expansion method. Without
loss of generality, we assume that $\bm Z_i$ satisfy the relation $\sum_{i=1}^{A}\bm Z_i=0$.

The AMD wave function given by Eq. \eqref{eq:amdwf1} is not an eigenstate of the parity and angular 
momentum. Therefore, the parity and angular momentum projections are usually performed. 
\begin{align}
 \Psi^{J\pi}_{MKA} &= \frac{1}{\sqrt{\mathcal{N}^{J\pi}_K}}
 \hat{P}^{J}_{MK} \hat{P}^\pi \Psi_A^{int},\label{eq:amdwf2}\\
 \mathcal{N}^{J\pi}_K &= \Braket{\Psi_{A}^{int}|\hat{P}^{J}_{KK}\hat{P}^\pi|\Psi_{A}^{int}},\\
 \hat{P}^{J}_{MK} &=\int d\Omega D^{J*}_{MK}(\Omega)\hat{R}(\Omega),\quad
 \hat{P}^{\pi}=\frac{1+\pi\hat{P}_r}{2}, \quad \pi=\pm,
\end{align}
where $\hat{P}^{J}_{MK}$ and $\hat{P}^\pi$ are the angular momentum and parity projectors.
$D^{J}_{MK}(\Omega)$ and $\hat{R}(\Omega)$ are the Wigner $D$ function and rotation operator
dependent on the Euler angles $\Omega$. $\hat{P}_r$ is the parity operator.

In addition to the projection, in nuclear structure studies, the parity and angular momentum
projected AMD wave functions are superposed to take the effects of the configuration mixing
and the shape fluctuation into account (generator coordinate method; GCM). 
\begin{align}
 \Psi^{J\pi}_{MA} &= \sum_{s=1}^{s_{max}}\sum_{K=-J}^Jc_{sK}
 \frac{1}{\sqrt{\mathcal N_{K}^{J\pi}(s)}}\hat{P}^{J}_{MK}\hat{P}^\pi\Psi_A^{int}(s)=
 \sum_{s=1}^{s_{max}}\sum_{K=-J}^Jc_{sK}\Psi_{MKA}^{J\pi}(s),\label{eq:amdwf3}\\
 \Psi_{MKA}^{J\pi} (s)&= \frac{1}{\sqrt{\mathcal N_{K}^{J\pi}(s)}}
 \hat{P}^{J}_{MK}\hat{P}^{\pi}\Psi_{A}^{int}(s),\\
 \mathcal N^{J\pi}_{K}(s)
 &=\Braket{\Psi_{A}^{int}(s)|\hat{P}^{J}_{KK}\hat{P}^\pi|\Psi_{A}^{int}(s)},
\end{align}
where $s$ is the index for the internal wave functions and $c_{sK}$ is the coefficient of the
superposition. Hereafter, we denote the wave functions given by Eq. \eqref{eq:amdwf1}, 
\eqref{eq:amdwf2} and \eqref{eq:amdwf3} as ``AMD wave function'', ``projected AMD wave function''
and ``GCM wave function'', respectively.

It must be noted that the following discussion and the Laplace expansion method are also
applicable to the Brink-Bloch wave function \cite {Brink1966} and $SU(3)$ shell model wave
function 
(harmonic oscillator wave function without the spin-orbit splitting), because the AMD wave 
function includes the Brink-Bloch wave function and $SU(3)$ shell model wave function 
as its spacial cases. When
the centroid of the wave packets are common for the quartet of $n\uparrow$, $n\downarrow$,
$p\uparrow$ and $p\downarrow$, the AMD wave function is equal to the Brink-Bloch wave function for
$N\alpha$ systems. At the limit of the $\bm Z_i \rightarrow 0$ the AMD wave function is equal to
the $SU(3)$ shell model wave function.

\subsection{Laplace expansion of the AMD wave function}
The Laplace expansion of the determinant of an $A\times A$ matrix $B$ is given as,
\begin{align}
  |B| = \sum_{1\leq i_1<\dots<i_{C_1}\leq A} P(i_1,\dots,i_{C_1}) 
  |B(i_1,\dots, i_{C_1})| |{B}(i_{C_1+1},\dots, i_{A})|,
  \end{align}
where the summation runs over all possible combinations of indices $i_1,\dots, i_{C_1}$.
The phase factor $P(i_1,\dots, i_{C_1})$ is defined as
\begin{align}
  P(i_1,\dots,i_{C_1}) = (-)^{C_1(C_1+1)/2+ \sum_{s=1}^{C_1} i_s}.
\end{align}
$|B(i_1,\dots,i_{C_1})|$ is the determinant of the $C_1 \times C_1$ matrix composed
from  the $1,...,C_1$th rows and the $i_1,\dots, i_{C_1}$th columns of the matrix $B$ 
\begin{align}
 |B(i_1,\dots,i_{C_1})| = \left|
 \begin{matrix}
  B_{1i_1}      & \dots  & B_{1i_{C_1}} \\
  \vdots& \ddots & \vdots\\
  B_{C_1i_1} & \dots  & B_{C_1 i_{C_1}}
  \end{matrix}
  \right|,
\end{align}
and $|{B}(i_{C_1+1},\dots i_{A})|$ is the determinant of the $C_2 \times C_2$
matrix ($C_1+C_2=A$) formed by removing the $1,\dots,C_1$th rows and the $i_1,\dots,i_{C_1}$th
columns from $B$,
\begin{align}
 |{B}(i_{C_1+1},\dots, i_{A})| = \left|
  \begin{matrix}
    B_{C_1+1,i_{C_1+1}}  & \dots  & B_{C_1+1,i_{A}} \\
    \vdots               & \ddots & \vdots            \\
    B_{A,  i_{C_1+1}}        & \dots  & B_{A,  i_{A}}
  \end{matrix}
  \right|,
\end{align}
where $i_{C_1+1},...,i_A$ denote the column indices other than $i_1,...,i_{C_1}$ and satisfy the
relation $1\leq i_{C_1+1} < ... < i_{A}\leq A$. 

Applying the Laplace expansion to the $A$-body AMD wave function given by Eq. (\ref{eq:amdwf1}),
we obtain the decomposition of the AMD wave function,
\begin{align}
 \Psi_{A}^{AMD} = \sqrt{\frac{C_1!C_2!}{A!}}
 \sum_{1\leq i_1<\dots<i_{C_1}\leq A} P(i_1,\dots, i_{C_1}) 
 \Psi_{C_1}^{AMD}(i_ 1,\dots, i_{C_1})
 \Psi_{C_2}^{AMD}({i_{C_1+1},\dots,i_{A}}).\label{eq:laplace1}
\end{align}
Here, $\Psi_{C_1}^{AMD}(i_ 1,\dots,i_{C_1})$ and ${\Psi}_{C_2}^{AMD}({i_{C_1+1},\dots, i_{A}})$ are
the AMD wave functions for the subsystems with masses $C_1$ and $C_2$ which are defined as  
\begin{align}
  \Psi_{C_1}^{AMD}(i_1,\dots, i_{C_1}) &= \frac{1}{\sqrt{C_1!}} 
  \left| \begin{matrix}
    \braket{\bm{r}_1|\psi_{i_1}} &  \dots & \braket{\bm{r}_1|\psi_{i_{C_1}}} \\
    \vdots &  \ddots & \vdots\\
    \braket{\bm{r}_{C_1}|\psi_{i_1}} & \dots & \braket{\bm{r}_{C_1}|\psi_{i_{C_1}}} 
  \end{matrix}\right|,\\
 \Psi_{C_2}^{AMD}({i_{C_1+1},,\dots,i_{A}}) &= \frac{1}{\sqrt{C_2!}} 
  \left| \begin{matrix}
    \braket{\bm{r}_{C_1+1}|\psi_{i_{C_1+1}}} &  \dots & \braket{\bm{r}_{C_1+1}|\psi_{i_{A}}} \\
    \vdots &  \ddots & \vdots\\
    \braket{\bm{r}_{A}|\psi_{i_{C_1+1}}} & \dots & \braket{\bm{r}_{A}|\psi_{i_{A}}} \\
  \end{matrix}\right|.
\end{align}
Since the internal and the center-of-mass wave functions are analytically separable as shown in 
Eq. (\ref{eq:cm1}), the product of the AMD wave functions in the right hand side of
Eq. (\ref{eq:laplace1}) is equal to a product of the internal and center-of-mass wave functions 
of the subsystems,  
\begin{align}
 \Psi_{C_1}^{AMD}\Psi_{C_2}^{AMD} &= \Psi_{C_1}^{cm}\Psi_{C_2}^{cm}\Psi_{C_1}^{int}\Psi_{C_2}^{int}, \\
 \Psi_{C_1}^{cm} &= \left(\frac{|2C_1M|}{\pi^3}\right)^{1/4}
 \exp\left\{-C_1(\bm R_{C_1}- \bm Z_{C_1})^TM(\bm R_{C_1}- \bm Z_{C_1})\right\},\\
 \Psi_{C_2}^{cm} &= \left(\frac{|2C_2M|}{\pi^3}\right)^{1/4}
 \exp\left\{-C_2(\bm R_{C_2}- \bm Z_{C_2})^TM(\bm R_{C_2}- \bm Z_{C_2})\right\},\\
 \bm R_{C_1} &= \frac{1}{C_1}\sum_{i=1}^{C_1}\bm r_i, \quad
 \bm R_{C_2} = \frac{1}{C_1}\sum_{i=C_1+1}^{A}\bm r_i, \label{eq:rc}\\
 \bm Z_{C_1} &= \frac{1}{C_1}\sum_{i\in\{i_1,...,i_{C_1}\}}\bm Z_i,\quad
 \bm Z_{C_2} = \frac{1}{C_2}\sum_{i\in\{i_{C_1+1},...,i_A\}}\bm Z_i,
\end{align}
where we suppressed the indices $i_1,...,i_A$ for simplicity. $\bm R_{C_1}$ and $\bm R_{C_2}$
denote the center-of-mass coordinates of the subsystems. Then, we rewrite the product of the
center-of-mass wave functions of clusters to the product of the center-of-mass wave function of
$A$-body system $\Psi^{cm}_A$ and the relative wave function between the subsystems $\chi(\bm r)$.  
\begin{align}
 \Psi_{C_1}^{cm} \Psi_{C_2}^{cm} &= \Psi_A^{cm} \chi(\bm{r}),\label{eq:cmdcmp1}\\
 \chi(\bm{r}) &= \left(\frac{|2\Gamma|}{\pi^3}\right)^{1/4}
 \exp\left\{-(\bm{r}-\bm{z})^T\Gamma(\bm{r}-\bm{z})\right\},\\
 \bm r &= \bm R_{C_1} - \bm R_{C_2}, \quad  \bm{z} = \bm{Z}_{C_1} - \bm{Z}_{C_2}, \quad
 \Gamma = \frac{C_1C_2}{A}M. \label{eq:rc2}
\end{align}
As a result, the product of the AMD wave functions is transformed as follows, 
\begin{align}
 \Psi^{AMD}_{C_1}(i_1,...,i_{C_1})\Psi^{AMD}_{C_2}(i_{C_1+1},...,i_A) =
 \Psi^{cm}_A\chi(\bm{r};i_1,...,i_A)
 \Psi^{int}_{C_1}(i_1,...,i_{C_1})\Psi^{int}_{C_2}(i_{C_1+1},...,i_A).\label{eq:laplace2}
\end{align}
Note that  $\Psi^{cm}_A$ is independent of the choice of $i_1,\dots, i_{C_1}$. Substituting
Eq. \eqref{eq:laplace2} into Eq.~\eqref{eq:laplace1}, and removing the center-of-mass wave
function,  we obtain a decomposition of $A$-body internal wave function  into two subsystems with
masses $C_1$ and $C_2$.
\begin{align}
 \Psi^{int}_A = \sqrt{\frac{C_1!C_2!}{A!}}
 \sum_{1\leq i_1<\dots<i_{C_1}\leq A} P(i_1,\dots,i_{C_1}) 
 \chi(\bm{r};i_1, \dots i_{A})
 \Psi^{int}_{C_1}(i_ 1,\dots,i_{C_1})
 \Psi^{int}_{C_2}(i_{C_1+1},\dots, i_{A}). \label{eq:laplace3}
\end{align}
It is noted that the Laplace expansion can be applied recursively, and hence, the decompositions
of the A-body wave function into three and more subsystems are also straightforward.

\subsection{Calculation of the RWA using Laplace expansion}
Using Laplace expansion of AMD wave function, we can calculate RWA of the $C_1+C_2$ cluster system
without any approximation. First, we discuss the RWA of a single projected AMD wave function, and
the extension to the GCM wave function is discussed later. 

The RWA for two-body cluster system is defined as the overlap amplitude between the $A$-body wave
function $\Psi^{J\pi}_{MA}$ and the reference state composed of the clusters with masses $C_1$ and
$C_2$, 
\begin{align} 
  y^{J\pi}_{j_1{\pi_1}j_2{\pi_2}j_{12}l}(a) =& 
  \sqrt{\frac{A!}{(1+\delta_{C_1C_2})C_1!C_2!}}\Braket{ \frac{\delta(r-a)}{r^2}
  \left[Y_{l}(\hat{r}) \left[\Phi^{j_1{\pi_1}}_{C_1} \Phi^{j_2{\pi_2}}_{C_2} \right]_{j_{12}}
  \right]_{JM} | \Psi^{J\pi}_{MA}},\label{eq:rwa1}
\end{align}
where $\Phi^{j_1\pi_1}_{C_1}$ and $\Phi^{j_2\pi_2}_{C_2}$ are the wave functions of clusters $C_1$
and $C_2$. Their spins $j_1$ and  $j_2$ are coupled to $j_{12}$, and
$j_{12}$ is coupled to orbital angular momentum $l$ of the inter-cluster motion to yield the total 
spin-parity $J^\pi$. Therefore, $\pi_1$, $\pi_2$ and $l$ must satisfy the relation
$\pi=\pi_1\pi_2(-)^l$. We assume that the wave functions $\Psi^{J\pi}_{MA}$,
$\Phi_{C_1}^{j_1\pi_1}$ and $\Phi_{C_2}^{j_2\pi_2}$ are antisymmetrized and normalized. 

With this definition, by substituting Eq. \eqref{eq:amdwf2} into \eqref{eq:rwa1}, the RWA of a
projected AMD wave function reads, 
\begin{align} 
  y^{J\pi}_{j_1{\pi_1}j_2{\pi_2}j_{12}l}(a) =& 
  \sqrt{\frac{A!}{(1+\delta_{C_1C_2})C_1!C_2!}}\Braket{\frac{\delta(r-a)}{r^2}
  \left[Y_{l}(\hat{r})\left[\Phi^{j_1{\pi_1}}_{C_1} \Phi^{j_2{\pi_2}}_{C_2}
  \right]_{j_{12}} \right]_{JM} | \Psi^{J\pi}_{MKA}}\nonumber\\
  =&  \frac{1}{\sqrt{\mathcal{N}^{J\pi}_K}} \sqrt{\frac{A!}{(1+\delta_{C_1C_2})C_1!C_2!}}
 \Braket{ \frac{\delta(r-a)}{r^2}\hat{P}^{J}_{KM}\left[Y_{l}(\hat{r})
 \left[\Phi_{C_1}^{j_1\pi_1} \Phi_{C_2}^{j_2\pi_2} \right]_{j_{12}}  
 \right]_{JM} | \Psi_A^{int}}\nonumber\\
  =&  \frac{1}{\sqrt{\mathcal{N}^{J\pi}_K}} \sqrt{\frac{A!}{(1+\delta_{C_1C_2})C_1!C_2!}}
 \Braket{ \frac{\delta(r-a)}{r^2}\left[Y_{l}(\hat{r})
 \left[\Phi_{C_1}^{j_1\pi_1} \Phi_{C_2}^{j_2\pi_2} \right]_{j_{12}} 
 \right]_{JK} | \Psi_A^{int}}\nonumber\\
  =& \frac{1}{\sqrt{\mathcal{N}^{J\pi}_K}} \sqrt{\frac{A!}{(1+\delta_{C_1C_2})C_1!C_2!}}
 \sum_{m_{12}m_lm_1m_2} C^{JK}_{lm_lj_{12}m_{12}} C^{j_{12}m_{12}}_{j_1m_1j_2m_2}\nonumber\\
 &\times\Braket{ \frac{\delta(r-a)}{r^2}Y_{lm_l}(\hat{r}) 
 \Phi_{m_1C_1}^{j_1\pi_1} \Phi_{m_2C_2}^{j_2\pi_2} | \Psi_{A}^{int}},\label{eq:rwa2}
\end{align}
where we used the relation $\pi=\pi_1\pi_2(-)^l$ and the properties of the angular momentum
projector $(\hat{P}^{J}_{MK})^\dagger = \hat{P}_{KM}^J$ and 
$\hat{P}^{J\pi}_{KM}\ket{JM} = \ket{JK}$. $C^{JM}_{j_1m_1j_2m_2}$ denotes the Clebsch-Gordan
coefficient. By using the Laplace expansion of $\Psi_{A}^{int}$ given by Eq. \eqref{eq:laplace3},
the  braket in the right hand side of  Eq. (\ref{eq:rwa2}) is written as
\begin{align}
 &\Braket{ \frac{\delta(r-a)}{r^2}Y_{lm_l}(\hat{r})
 \Phi^{j_1\pi_1}_{m_1C_1} \Phi^{j_2\pi_2}_{m_2C_2}| \Psi_{A}^{int}} 
 = \sqrt{\frac{C_1!C_2!}{A!}}\sum_{1\leq i_1<...<i_{C_1} \leq A} P(i_1, \dots, i_{C_1})\nonumber\\ 
 &\times \Braket{\frac{\delta(r-a)}{r^2}Y_{lm_l}(\hat{r})
 \Phi^{j_1\pi_1}_{m_1C_1} \Phi^{j_2\pi_2}_{m_2C_2}  | \chi(\bm{r};i_1, \dots,i_{A})
 \Psi_{C_1}^{int}(i_1,\dots,i_{C_1})\Psi_{C_2}^{int}(i_{C_1+1}, \dots, i_{A})
 }.\label{eq:rwa3}
\end{align}
Note that the braket in the last line has no antisymmetrizer with respect to the nucleons
belonging to different subsystems. Therefore, it is equal to the product of the overlaps between 
the relative wave functions and between the subsystems.
\begin{align}
 &\Braket{   \frac{\delta(r-a)}{r^2}Y_{lm_l}(\hat{r}) 
 \Phi^{j_1{\pi_1}}_{m_1C_1} \Phi^{j_2\pi_2}_{m_2C_2}|\chi(\bm{r};i_1, \dots,i_{A})
  \Psi_{C_1}^{int}(i_1,\dots,i_{C_1})\Psi_{C_2}^{int}(i_{C_1+1}, \dots, i_{A})
 }\nonumber\\
 &= \Braket{\frac{\delta(r-a)}{r^2}Y_{lm_l}(\hat{r})|\chi(\bm{r};i_1,\dots,i_{A})}
\braket{\Phi^{j_1\pi_1}_{m_1C_1}|\Psi_{C_1}^{int}(i_1,\dots,i_{C_1})}
 \braket{\Phi^{j_2{\pi_2}}_{m_2C_2}|\Psi_{C_2}^{int}(i_{C_1+1},\dots,i_{A})}.
 \label{eq:rwa4}
\end{align}
Substituting Eqs. \eqref{eq:rwa3} and \eqref{eq:rwa4} to Eq. \eqref{eq:rwa2}, we obtain the RWA of
a projected AMD wave function. 
\begin{align}
  &y^{J\pi}_{j_1{\pi_1}j_2{\pi_2}j_{12}l}(a) = 
  \frac{1}{\sqrt{\mathcal{N}^{J\pi}_K(1+\delta_{C_1C_2})}}
  \sum_{1\leq i_1<...<i_A\leq A} P(i_1,\dots,i_{C_1}) \nonumber\\
 &\times\left[  \chi_{l}(a;i_1,\dots, i_{A})\left[
 N^{j_1\pi_1}(i_1,\dots,i_{C_1})N^{j_2\pi_2}(i_{C_1+1},\dots i_{A})
  \right]_{j_{12}}\right]_{JK},\label{eq:rwa5}
\end{align}
with the definitions of the overlaps,
\begin{align}
 \chi_{lm_l}(a;i_1,\dots,i_{A}) &= 
 \Braket{\frac{\delta(r-a)}{r^2}Y_{lm_l}(\hat{r})|\chi(\bm{r};i_1,\dots,i_{A})},\label{eq:ov1}\\
 N^{j_1\pi_1}_{m_1}(i_1,\dots,i_{C_1}) &= 
 \braket{\Phi^{j_1\pi_1}_{m_1C_1}|\Psi_{C_1}^{int}(i_1,\dots,i_{C_1})},\label{eq:ov2}\\
 N^{j_2\pi_2}_{m_2}(i_{C1+1},\dots,i_{A}) &= 
 \braket{\Phi^{j_2{\pi_2}}_{m_2C_2}|\Psi_{C_2}^{int}(i_{C_1+1},\dots,i_{A})}.\label{eq:ov3}
\end{align}
Thus, the RWA is obtained by calculating the overlaps defined by Eqs. \eqref{eq:ov1},
\eqref{eq:ov2} and \eqref{eq:ov3} which are easily calculated as explained in the appendix
\ref{app:appa}.

The extension of the method to the GCM wave function is straightforward. Substituting
Eq. \eqref{eq:amdwf3} to Eq. \eqref{eq:rwa1}, one easily obtains the RWA of GCM wave function as   
follows.  
\begin{align}
 y^{J\pi}_{j_1{\pi_1}j_2{\pi_2}j_{12}l}(a) &= \sum_{s=1}^{s_{max}}\sum_{K=-J}^J c_{sK}
 y^{J\pi}_{j_1{\pi_1}j_2{\pi_2}j_{12}l}(a;sK) \\
 y^{J\pi}_{j_1{\pi_1}j_2{\pi_2}j_{12}l}(a;sK) &=
 \sqrt{\frac{A!}{C_1!C_2!}}\Braket{  \frac{\delta(r-a)}{r^2}
 \left[Y_{l}(\hat{r})\left[\Phi^{j_1{\pi_1}}_{C_1}\Phi^{j_2{\pi_2}}_{C_2}\right]_{j_{12}}
 \right]_{JM} | \Psi^{J\pi}_{MKA}(s)}.\label{eq:rwa6}
\end{align}
Thus, the RWA of the GCM wave function is the superposition of the RWAs of the projected AMD wave  
functions defined by Eq. \eqref{eq:rwa6} which are calculated by using Eq. \eqref{eq:rwa5} for
every $\Psi^{J\pi}_{MKA}(s)$. In a same way, when the reference wave function
$\Phi_{m_1C_1}^{j_1\pi_1}$  is a GCM wave function, the overlap
$N^{j_1\pi_1}_{m_1}(i_1,...,i_{C_1})$ is a sum of the overlaps of the projected AMD wave
functions. 

\subsection{Advantages of the Laplace expansion method}
It may be worthwhile to compare the Laplace expansion method with an ordinary method 
\cite{Horiuchi1972,Horiuchi1977,Kanada-Enyo2003a,Kimura2004a} which is often used in the cluster
models and AMD to see its advantages and disadvantages.  The ordinary method uses a set of the
projected Brink-Bloch type wave functions defined  as  
\begin{align}
 \Phi^{J\pi}_{j_1\pi_1j_2\pi_2j_{12}l}(S_p) &=\frac{2l+1}{2J+1}
 \sum_{m_{12}m_1m_2}C^{Jm_{12}}_{l0j_{12}m_{12}}\hat{P}^{J}_{Mm_{12}}
 C^{j_{12}m_{12}}_{j_1m_1j_2m_2}\Phi^{BB}_{j_1\pi_1m_1j_2\pi_2m_2}(S_p),\label{eq:wfbb1}\\
 \Phi^{BB}_{j_1\pi_1m_1j_2\pi_2m_2}(S_p) &=\sqrt{\frac{C_1!C_2!}{A!}}\mathcal A
 \left\{\hat{P}^{j_1}_{m_1k_1}\hat{P}^{\pi_1} \Phi_{C_1}\left(-\frac{C_2}{A}\bm S_p\right) 
 \hat{P}^{j_2}_{m_2k_2}\hat{P}^{\pi_2} \Phi_{C_2}\left(\frac{C_1}{A}\bm S_p\right)\right\},
 \label{eq:wfbb2}\\
 \bm S_p &= (0,0,S_p).
\end{align}
$\Phi_{C_1}\left(-{C_2}/{A}\bm S_p\right)$ and $\Phi_{C_2}\left({C_1}/{A}\bm S_p\right)$
are the wave functions for clusters with masses $C_1$ and $C_2$ with their center-of-mass wave
functions, and placed with the inter-cluster distance $S_p$. The inter-cluster distance $S_p$ is
discretized, for example, as 
\begin{align}
 S_p = p\Delta S, \quad p=1,...,p_{max}. 
\end{align}
The following three conditions are often required to reduce the computational cost. 
\begin{itemize}
 \item $\Phi_{C_1}$ and $\Phi_{C_2}$ are the $SU(3)$ shell model wave functions without
       particle-hole excitations. 
 \item The oscillator parameters of $\Phi_{C_1}$ and $\Phi_{C_2}$ are the same value, 
       $\hbar\omega=2\hbar^2\nu/m$
 \item $\Phi_{C_1}$ and $\Phi_{C_2}$ are the eigenstates of the principal quantum number $\hat{N}$
\end{itemize}
With these conditions are satisfied, the RWA is given as follows,
\begin{align}
 &y^{J\pi}_{j_1\pi_1j_2\pi_2j_{12}l}(a) = \frac{1}{\sqrt{1+\delta_{C_1C_2}}}\sum_{N} \mu_{Nl} e_N
 R_{Nl}(r), \label{eq:ord1}\\ 
 &\mu_{Nl} = \Braket{ R_{Nl}(r)
  \left[ Y_l(\hat r)\left[\Phi^{j_1{\pi_1}}_{C_1} \Phi^{j_2{\pi_2}}_{C_2} \right]_{j_{12}}
  \right]_{J} | \mathcal A
 \left\{R_{Nl}(r)\left[ Y_l(\hat r)\left[\Phi^{j_1{\pi_1}}_{C_1}
 \Phi^{j_2{\pi_2}}_{C_2} \right]_{j_{12}} \right]_{J} \right\}},\label{eq:ord2}\\
 &e_{Nl} =(-)^{(N-l)/2}\sqrt{\frac{(2l+1)}{(N-l)!!(N+l+1)!!}}
 \sum_{pq}\frac{(\nu S_p^2)^{N/2}}{\sqrt{N!}}e^{-\nu S_p^2/2}
 B_{pq}^{-1}\braket{\Phi^{J\pi}_{j_1\pi_1j_2\pi_2j_{12}l}(S_q)|\Psi^{J\pi}_{MA}}, 
 \label{eq:ord3}\\
 &B_{pq} = \braket{\Phi^{J\pi}_{j_1\pi_1j_2\pi_2j_{12}l}(S_p)
 |\Phi^{J\pi}_{j_1\pi_1j_2\pi_2j_{12}l}(S_q)}.\label{eq:ord4}
\end{align}
Here $R_{Nl}(r)$ is the radial wave function of harmonic oscillator (HO). 
The derivation of these equations is explained in the appendix \ref{app:appb} and the
Refs. \cite{Horiuchi1972,Horiuchi1977,Kanada-Enyo2003a,Kimura2004a}. From these equations, we can
see several advantages of the Laplace expansion  method listed below. 

\begin{enumerate}
 \item The size of the Gaussian wave packets describing clusters $C_1$ and $C_2$ in the reference
       state can be different in the Laplace expansion method. This is an advantage when we
       calculate the RWA of the different size clusters such as $^{16}{\rm O}+\alpha$ and
       $^{40}{\rm Ca}+\alpha$ \cite{Tohsaki-Suzuki1978}.  \\
       On the other hand, in the ordinary method, 
       they must be equal to analytically separate the center-of-mass and relative wave functions
       as explained in the appendix \ref{app:appb}. 
 \item The deformed Gaussian wave packets can be used to describe the clusters $C_1$ and $C_2$ in
       the reference state. Therefore, the Laplace expansion method can easily calculate the RWA
       of deformed clusters such as $^{10}{\rm Be}+\alpha$ and $^{24}{\rm Mg}+\alpha$ without any
       approximation. \\ 
       On the other hand, for the analytical separation of the center-of-mass and the relative
       wave functions, ordinary method uses spherical Gaussian. For example, in Ref. 
       \cite{Baba2016,Baba2017},  $^{10}{\rm Be}$ wave function was approximated by the spherical
       Gaussian to estimate the $^{10}{\rm Be}+\alpha$ RWA. 
 \item The angular momentum projection of the cluster wave functions can be done with much reduced
       computational cost. This is another advantage when we calculate the RWA of the deformed 
       clusters.  \\ 
       When the intrinsic wave function of clusters $\Phi_{C_1}$ and $\Phi_{C_2}$ are not the
       eigenstate of the angular momentum, we need to perform the angular momentum projection of
       each cluster. In the ordinary method, we need to calculate Eqs. \eqref{eq:ord3} and
       \eqref{eq:ord4}, which contain  three and five angular momentum projectors and demand huge
       computational cost. Therefore, the approximation is often applied
       \cite{Baba2016,Baba2017}. However, in the case of the Laplace expansion method, we only
       need to calculate Eqs. \eqref{eq:ov2} and  \eqref{eq:ov3} which contain only one angular
       momentum projector. 
 \item The GCM wave function can be used as the wave functions of clusters. Therefore, the Laplace
       expansion method can treat various clusters which cannot be described by a single AMD wave
       function. A typical example is $^{6}{\rm He}$ cluster which has neutron halo. In the
       ordinary method, the $^{6}{\rm He}$ cluster is often approximated by the
       $(0s)^4(0p_{3/2})^2$ configuration of HO wave function \cite{Kanada-Enyo2003a,Baba2017}.  
 \item Laplace expansion method does not use the eigenvalue of norm kernel defined by
       Eq. \eqref{eq:ord2}, which represents the antisymmetrization effect between clusters. In
       general, the calculation of this quantity is not easy when the clusters are not described
       by the $SU(3)$ shell model  wave functions. Therefore, several approximations have been
       suggested and applied \cite{Kanada-Enyo2003a,Kanada-Enyo2014b}. Laplace expansion method is
       free from such approximations.  
\end{enumerate}
The disadvantage of Laplace expansion method should be also commented. It is clear from
Eq. \eqref{eq:rwa5}, that the computational cost greatly increases when the mass of the system $A$ is
large and the masses of clusters are equal ($C_1=C_2$), because the number of possible
combinations of $i_1,...,i_A$ becomes huge. A typical example is the $^{16}{\rm O}+{}^{16}{\rm O}$ 
cluster in $^{32}{\rm S}$. In this case, there are 
$\left({16!}/({8!8!})\right)^2\simeq 1.6\times 10^8$ combinations of $i_1,...,i_A$. On the other
hand, since $^{16}{\rm O}$ is a spherical and $SU(3)$ scalar cluster, the ordinary method can be
straightforwardly applied and quickly calculated \cite{Kimura2004}.

\section{Numerical examples}\label{sec:result}
In this section, we present the numerical results of the RWAs of the $^{16}{\rm O}+\alpha$ and
$^{24}{\rm Mg}+\alpha$ clustering in $^{20}{\rm Ne}$ and $^{28}{\rm Si}$ which are composed of the
unequal size clusters and deformed cluster. The wave functions of $^{20}{\rm Ne}$ and 
$^{28}{\rm Si}$ are calculated by the antisymmetrized molecular dynamics and the same with those
obtained in our previous studies \cite{Kimura2004a,Chiba2016,Taniguchi2009,Chiba2016a}. The
Hamiltonian is common to both nuclei and is given as, 
\begin{align}
 \hat{H} = \sum_{i=1}^A\hat{t}_i - \hat{t}_{cm} + \sum_{i<j}^A\hat{v}_{NN}
 + \sum_{i<j}^A\hat{v}_{Coul},
\end{align}
where $\hat{t}_i$ and $\hat{t}_{cm}$ denote the nucleon and the center-of-mass kinetic energies.  
$\hat{v}_{NN}$ and $\hat{v}_{Coul}$ denote the Gogny D1S effective nucleon-nucleon interaction
\cite{Berger1991} and the Coulomb interaction. The detailed set up of the calculations is explained
below. 

\subsection{RWA of the $^{16}{\rm O}+\rm \alpha$ clustering in $^{20}{\rm Ne}$}
The $^{16}{\rm O}+\alpha$ clustering of $^{20}{\rm Ne}$ is very famous and experimentally
identified well \cite{Fujiwara1980}. There are three rotational bands with 
$^{16}{\rm O}+\alpha$ cluster structure, which are built on the $0^+_1$, $1^-_1$ (5.8 MeV)  and
$0^+_4$ (8.7 MeV) states, respectively. Here we discuss the RWAs of the $0^+_1$, $1^-_1$  and
$0^+_4$ states as an example of the unequal size clusters.  

\begin{figure}[h]
 \centering
 \includegraphics[width=0.6\hsize]{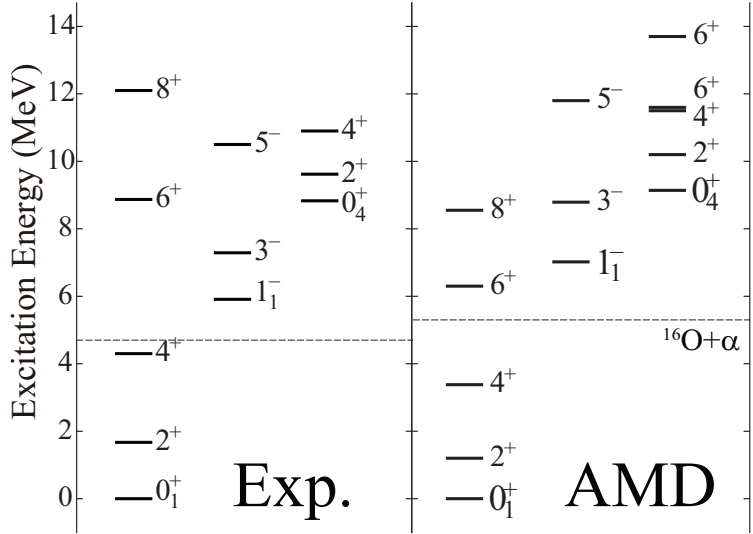}
 \caption{The calculated and observed partial level scheme of $^{20}{\rm Ne}$ corresponding
 to the $^{16}{\rm O}+\alpha$ cluster states. The dashed horizontal lines indicate experimental 
 and theoretical $^{16}{\rm O}+\alpha$ cluster threshold energies.}
 \label{fig:level_ne20}
\end{figure}

The AMD wave functions of $^{20}$Ne for GCM calculation are prepared by the energy variation
with the constraint on the nuclear quadrupole deformation parameter $\beta$. The value of
parameter $\beta$ is constrained from 0.0 to 0.85 with an interval of 0.05. In addition to this, 
we also included Brink-Bloch type wave functions with the inter-cluster distance $S_p$
ranging from 1.0 fm to 8.0 fm with an interval of 1.0 fm. Here, the $^{16}{\rm O}$ cluster is 
described by a single AMD wave function obtained by the energy variation, while $\alpha$ cluster is
assumed to have $(0s)^4$ configuration. To analytically remove the center-of-mass motion from GCM
wave function, both clusters are assumed to have the same spherical oscillator parameter
$\nu=0.16$ fm$^{-2}$. In short, we superposed 18 AMD wave functions and 8 Brink-Bloch type wave
functions together, and solved the GCM. Thus-obtained level scheme is shown in
Fig.~\ref{fig:level_ne20}   together with the observed levels. The detailed discussions of these
states are found in the Refs. \cite{Kimura2004a,Chiba2016}.  

\begin{figure}[h]
 \centering
 \includegraphics[width=\hsize]{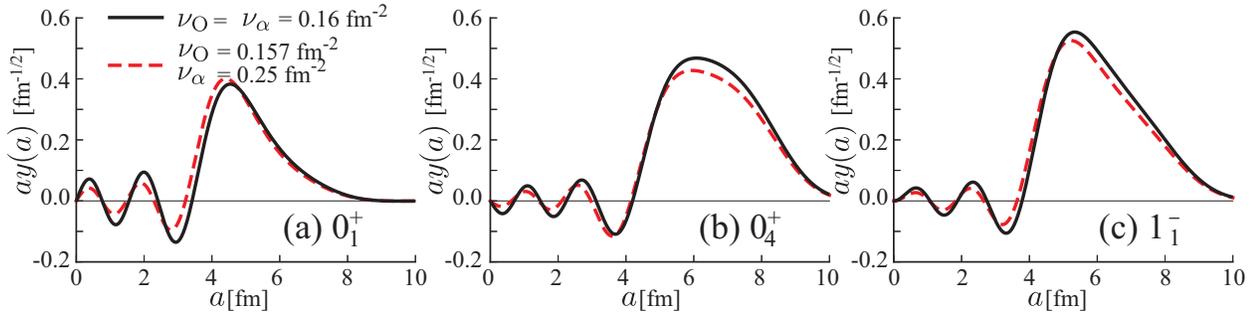}
 \caption{The $^{16}{\rm O}+\alpha$ cluster RWAs of  (a) the $0^+_1$ state (b) the $0^+_4$ state
 and (c) the $1^-_1$ state. The solid line show the RWAs obtained by using the reference states
 with  common oscillator lengths $\nu_{\rm O}=\nu_\alpha=0.16$ fm$^{-2}$, while the dashed lines
 show  those obtained by using the different oscillator parameters $\nu_{\rm O} = 0.157$ fm$^{-2}$
 and   $\nu_\alpha=0.25$ fm$^{-2}$.}   \label{fig:rwa_ne20}
\end{figure}

We prepared two different sets of the cluster wave functions as the reference states 
($\Phi^{j_1\pi_1}_{m_1C_1}$ and $\Phi^{j_2\pi_2}_{m_2C_2}$ in Eqs. \eqref{eq:ov2} and
\eqref{eq:ov3}) to evaluate RWA. In the first set,  $^{16}{\rm O}$ and $\alpha$ clusters have the
common oscillator lengths $\nu_{\rm O}=\nu_\alpha=0.16$ fm$^{-2}$. Namely they are same with the
above-mentioned Brink-Bloch type wave functions. In the second set, the oscillator lengths are
different. For $\alpha$ cluster, we used $\nu_\alpha = 0.25$ fm$^{-2}$, while we used 
$\nu_{\rm O}=0.157$ fm$^{-2}$ for $^{16}{\rm O}$ cluster which minimizes the intrinsic energy of
$^{16}{\rm O}$. In the following, we denote the calculation with the first and second sets of the
cluster wave functions by ``common size calculation'' and ``unequal size calculation'',
respectively. In both cases the RWAs were calculated by the Laplace expansion method.  The 
calculated RWAs for the $0^+_1$, $0^+_2$, and $1^-_1$ states are shown in 
Fig. \ref{fig:rwa_ne20}. And the $\alpha$ spectroscopic factor $S_\alpha$ and the dimensionless
decay width $\theta^2$ are listed in Table~\ref{tab:sfactor}, which are defined as follows.
\begin{align}
 S_\alpha &= \int^\infty_0 da \left| ay^{J\pi}_{j_1\pi_1j_2\pi_2j_{12}l}(a)\right|^2, \\
 \theta^2_\alpha &= \frac{a}{3} \left|ay^{J\pi}_{j_1\pi_1j_2\pi_2j_{12}l}(a)\right|^2.
\end{align}

\begin{table}[h]
 \caption{The $\alpha$ spectroscopic factor $S_\alpha$ and the dimension decay width
 $\theta^2_\alpha$  calculated at the channel radii  $a=6$ and 7 fm. } 
 \label{tab:sfactor}
 \centering
 \begin{tabular}{ccccccc}
  \hline
  & \multicolumn{3}{c}{$\nu_{\rm O}=\nu_\alpha=0.16$ fm$^{-2}$}&
  \multicolumn{3}{c}{$\nu_{\rm O}=0.157$ fm$^{-2}$, $\nu_\alpha=0.25$ fm$^{-2}$ }\\
  & $S_\alpha$    & $\theta^2_\alpha \ (a=6\ \rm{fm})$ & $\theta^2_\alpha \ (a=7\ \rm{fm})$ 
  & $S_\alpha$    & $\theta^2_\alpha \ (a=6\ \rm{fm})$ & $\theta^2_\alpha \ (a=7\ \rm{fm})$ \\
  \hline
  $0^+_1$  &  0.24         & 0.06         & 0.01
           &  0.26         & 0.05         & 0.01    \\
  $0^+_4$  &  0.62         & 0.44         & 0.43
           &  0.53         & 0.37         & 0.35   \\
  $1^-_1$  &  0.71         & 0.49         & 0.28
           &  0.63         & 0.41         & 0.22  \\
  \hline
 \end{tabular}
\end{table}

From larger amplitudes of the RWAs, $S_\alpha$ and $\theta_\alpha$, it is evident that the $0^+_4$
and $1^-_1$ states have more developed cluster structure than the ground state. When we compare
the RWAs obtained by the common size and unequal size calculations, we find the following
differences, although they show similar behavior. 

\begin{enumerate}
  \item The nodal points of RWA moves inward in the unequal size calculation.
  \item The amplitude of RWAs tend to be smaller  in the unequal size calculation.  
\end{enumerate}

The first point is due to the weaker antisymmetrization effect.  The unequal size calculation
uses much smaller size of the $\alpha$ cluster than the common size calculation. Therefore, the
$\alpha$ cluster is much less affected by the antisymmetrization effect. Since the oscillation of
the RWAs in the internal region originates in the antisymmetrization effect, the nodal positions of
RWA should move inward in the unequal calculation. 

For the second point, there may be two explanations. In the inner region, the $\alpha$ clusters
should be strongly distorted due to the strong effect of antisymmetrization and the mean-field
potential. In such case, the size of $\alpha$ cluster should differ from that of free $\alpha$
particle and may be enlarged to gain the attraction from the mean-field potential. Therefore, the
common size clusters may be favored in the inner region. 
In the outer region, the difference originates in the defect of the present GCM calculation. 
To analytically remove the center-of-mass wave function, the wave packet sizes of AMD wave
function are common to all nucleons. As a result, even at the large inter-cluster distance, the
oscillator parameters for $^{16}{\rm O}$ and $\alpha$ clusters are common in the GCM wave function,
while they should be unequal to describe the correct asymptotics. It is evident that the common
oscillator parameters of GCM wave function reduces the RWA in the outer region in the unequal size
calculation. From these differences, compared to the common size calculation, the unequal size
calculation tends to yield smaller values of $S_\alpha$ and $\theta_\alpha$ by approximately 10 to
20\% except for the ground state.

\subsection{RWA of the $^{24}{\rm Mg}+\alpha$ clustering in $^{28}{\rm Si}$}
A variety of cluster states such as the $^{24}{\rm Mg}+\alpha$, $^{20}{\rm Ne}+2\alpha$ and 
$^{16}{\rm O}+{}^{12}{\rm C}$ clustering are expected to exist in $^{28}{\rm Si}$. Many of them
are related to the nuclear reactions in the astrophysical processes, and hence, have been
intensively studied for many years 
\cite{Stokstad1972,Baye1976,Maas1978,Cseh1982,Tanabe1983,Kato1985,Kubono1986,Artemov1990,Ashwood2001,Shawcross2001,Taniguchi2009,Goasduff2014,Chiba2016a}, 
although their properties are not fully understood yet.  

In our resent study \cite{Chiba2016a}, we have performed the AMD calculation to identify these
cluster states. We made the energy variation with the constraint on the quadrupole deformation
parameters $\beta$ and $\gamma$ \cite{Kimura2012} to generate the AMD wave functions for the GCM
calculation.  In addition to this, we also made the energy variation with the constraint on the
inter-cluster distance \cite{Taniguchi2004} to generate various cluster configurations. These two
kinds of basis wave functions are superposed and the GCM calculation was performed. As a result,
we suggested various cluster bands; two groups of $^{24}{\rm Mg}+\alpha$ bands, the $^{20}{\rm
Ne}+2\alpha$ and   $^{16}{\rm O}+{}^{12}{\rm C}$ bands. Among them, we here discuss the RWAs of a
group of  $^{24}{\rm Mg}+\alpha$ bands, which was named  ``${}^{24}{\rm Mg}+\alpha$ (T) bands''
as an example of the deformed clusters.

\begin{figure}[h]
 \centering
 \includegraphics[width=0.95\hsize]{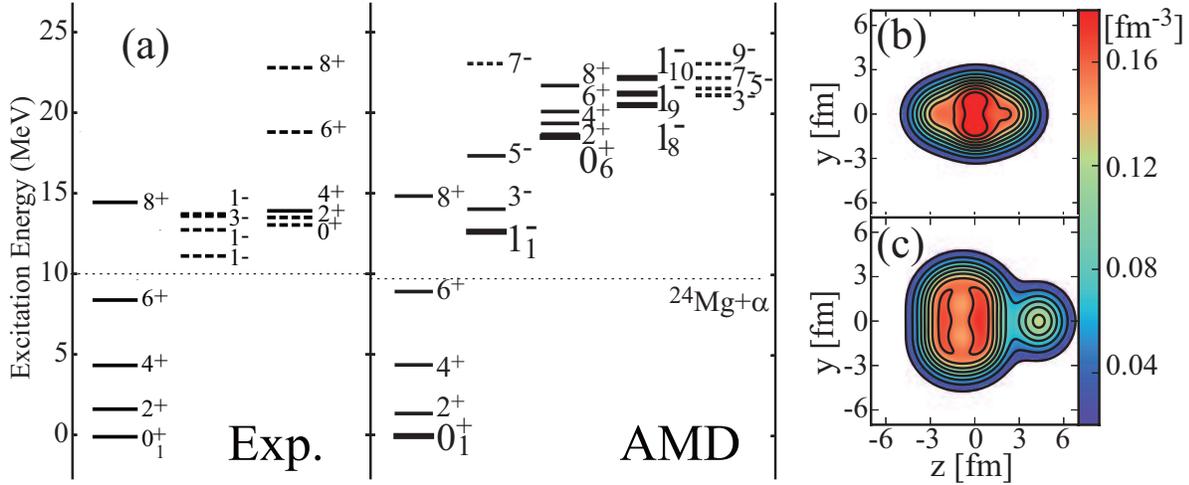}
 \caption{(a) The calculated and observed partial level scheme of $^{28}{\rm Si}$ corresponding
 to the $^{24}{\rm Mg}+\alpha$ cluster states. The dashed horizontal lines indicate experimental  
 and theoretical $^{24}{\rm Mg}+\alpha$ threshold energies. (b) and (c) Intrinsic density
 distributions of the AMD wave functions which have the maximum overlap with the $0^+_1$ and
 $0^+_6$ states, respectively.   \label{fig:level_si28}}
\end{figure}
Figure \ref{fig:level_si28} (a) shows the ${}^{24}{\rm Mg}+\alpha$ (T) bands. Three rotational
bands with pronounced ${}^{24}{\rm Mg}+\alpha$ clustering are built on the $1^-_1$ and $0^+_6$
states and on a group of $1^-$ states  ($1^-_8$, $1^-_9$ and $1^-_{10}$ states). Note that the
${}^{24}{\rm Mg}+\alpha$ configuration is strongly mixed with other cluster configurations such as
$^{16}{\rm O}+{}^{12}{\rm C}$. As a result, it does not appear as a single state in the band built
on the $1^-_{8,9,10}$ states. Therefore, in Fig. \ref{fig:level_si28} (a), we show the averaged
energy for the $3^-$, $5^-$, $7^-$ and $9^-$ states by the dotted lines. These three bands have
large overlap with the basis wave function shown in Fig. \ref{fig:level_si28} (c) in which the
longest axis of the deformed $^{24}{\rm Mg}$ cluster is perpendicular to the inter-cluster
coordinate between $^{24}{\rm Mg}$ and $\alpha$ clusters. The ground band is dominated by a
mean-field configuration shown in Fig. \ref{fig:level_si28} (b), but it also have non-negligible
overlap with the cluster configuration shown in Fig. \ref{fig:level_si28} (c). Therefore, the
ground band was also assigned as a member of ${}^{24}{\rm Mg}+\alpha$ (T) bands. Because
the $^{24}{\rm Mg}$ cluster is considerably deformed, we expect the rotational excitation of
$^{24}{\rm Mg}$ is coupled to the angular momentum of the inter-cluster motion in RWAs. 
Experimentally, the corresponding cluster bands are not clearly identified except for the ground
band. In Fig. \ref{fig:level_si28} we showed several candidates of the $^{24}{\rm Mg}+\alpha$
cluster states observed by the $\alpha$ transfer reaction on $^{24}{\rm Mg}$
\cite{Cseh1982,Artemov1990}.  

To calculate the RWAs of the $0^+_1$, $0^+_6$, $1^-_1$ and $1^-_{8,9,10}$ states, the cluster wave
functions in the reference state are prepared as follows.  The $\alpha$ cluster is assumed to have
a $(0s)^4$ configuration and its spherical oscillator parameter is set to $\nu_\alpha=0.25$
fm$^{-2}$. The AMD wave function for the $^{24}$Mg cluster is calculated by the energy variation,
and it is  projected to the $0^+$, $2^+$ and $4^+$ states. The oscillator parameter is determined
to minimize  the energy of the $0^+$ state. Because of the triaxial deformation of the 
$^{24}{\rm Mg}$, the optimum oscillator parameter is anisotropic and have different values for the
$x$, $y$ and $z$ directions as, $\bm \nu_{\rm Mg}=(0.12, 0.167, 0.169)$ fm$^{-2}$. 
Using these cluster wave functions, the RWA is calculated for various combinations of the angular
momenta. Denoting the parity and the angular-momentum of $^{24}$Mg by $j^\pi$ and those of the
inter-cluster motion by $l^{(-)^l}$, the RWAs of the $0^+$ states are calculated for the
combinations, $j^\pi \otimes l^{(-)^l} = 0^+ \otimes 0^+$, $2^+\otimes 2^+$ and $4^+ \otimes 4^+$,
and the RWAs of the $1^-$ states are calculated for  $j^\pi \otimes l^{(-)^l}=0^+ \otimes 1^-$,
$2^+ \otimes 1^-$, $2^+\otimes 3^-$, $4^+ \otimes 3^-$ and $4^+ \otimes 5^-$.

\begin{figure}[t]
  \centering
   \includegraphics[width=1.0\hsize]{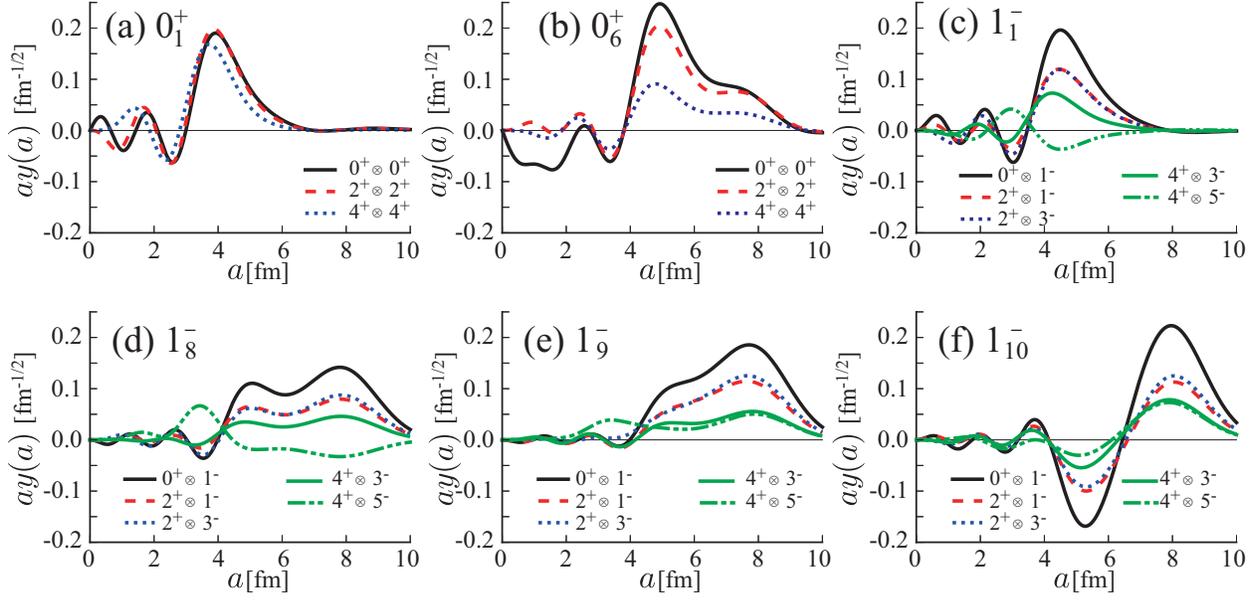}
  \caption{The RWAs of the $0^+_1$, $0^+_6$,
  $1^-_8$, $1^-_9$ and $1^-_{10}$ states of $^{28}$Si for $^{24}$Mg+$\alpha$ cluster channels 
  $j_2 \otimes l$ up to $j_2=4$ and $l=5$.}
  \label{fig:rwa_si28}
\end{figure}

\begin{table}[t]
 \caption{The calculated $\alpha$ spectroscopic factors $S_\alpha$ and the dimensionless decay
 widths $\theta^2_\alpha$ of  $^{24}$Mg+$\alpha$ clustering in $0^+$ states.  The dimensionless
 decay  widths are  given in the  unit of $10^{-2}$ and calculated with the channel radius $a=5$
 and 9 fm  for the  $0^+_1$ and  $0^+_6$ states, respectively.\label{tab:rwa_si28+}}   
  \centering
  \begin{tabular}{ccccccc}
   \hline
   &\multicolumn{3}{c}{$S_\alpha$}&\multicolumn{3}{c}{$\theta_\alpha^2$}\\
   $j^\pi \otimes l^{(-)^l}$& $0^+\otimes 0^+$ & $2^+\otimes 2^+$  & $4^+\otimes 4^+$ &
   $0^+\otimes 0^+$ & $2^+\otimes 2^+$  & $4^+\otimes 4^+$ \\
    \hline
    $0^+_1$     &  0.05     & 0.05      & 0.03  &  1.1 & 1.0 & 0.3  \\
    $0^+_6$     &  0.11     & 0.07      & 0.01  &  1.1 & 1.0 & 0.2  \\
   \hline
  \end{tabular}
\end{table}
\begin{table}[t]
 \caption{The calculated spectroscopic factors $S_\alpha$ and the dimensionless decay widths
 $\theta^2_\alpha$ of  $^{24}$Mg+$\alpha$ clustering in $1^-$ states.  The dimensionless decay
 widths are given in the unit of  $10^{-2}$ calculated with the channel radius $a=6$ and 9 fm for
 the $1^-_{1}$ and $1^-_{8,9,10}$ states, respectively.\label{tab:rwa_si28-}}  
 \centering
 \begin{tabular}{cccccc}
  \hline
  \multicolumn{6}{c}{$S_\alpha$}\\
  $j^\pi \otimes l^{(-)^l}$& $0^+\otimes 1^-$ & $2^+\otimes 1^-$ & $2^+\otimes 3^-$   
  & $4^+\otimes 3^-$ &  $4^+\otimes 5^-$  \\
  \hline
  $1^-_1$     &  0.06     & 0.02      & 0.02  & 0.03  & 0.01 \\
  $1^-_8$     &  0.06     & 0.02      & 0.04  & 0.01  & 0.01 \\
  $1^-_9$     &  0.09     & 0.04      & 0.02  & 0.01  & 0.01 \\
  $1^-_{10}$  &  0.12     & 0.03      & 0.04  & 0.01  & 0.01 \\
  \hline
  \multicolumn{6}{c}{$\theta_\alpha^2$}\\
  $j^\pi \otimes l^{(-)^l}$ & $0^+\otimes 1^-$ & $2^+\otimes 1^-$ & $2^+\otimes 3^-$   &
		  $4^+\otimes 3^-$  & $4^+\otimes 5^-$ \\
  \hline
  $1^-_1$     &  0.8      & 0.3      & 0.3  & 0.2 $\times 10^{-1}$ & 0.5 $\times 10^{-1}$ \\
  $1^-_8$     &  1.9      & 0.6      & 0.8  & 0.2 & 0.2  \\
  $1^-_9$     &  3.0      & 1.1      & 1.3  & 0.3 & 0.3  \\
  $1^-_{10}$  &  5.0      & 1.5      & 0.2  & 0.6 & 0.5  \\
  \hline
 \end{tabular}
\end{table}
The results are presented in Fig.~\ref{fig:rwa_si28}, 
Tabs. \ref{tab:rwa_si28+} and \ref{tab:rwa_si28-}. Although the detailed discussions on RWAs and
their relationship to the clustering in $^{28}{\rm Si}$ will be made in our next work, we here
briefly comment the characteristics of the calculated RWAs. 
The RWAs of the $0^+_1$ and $1^-$ states show similar nature. These states are dominated by the
mean-field configurations  and under the strong influence of the spin-orbit interaction. Therefore,
the clusters are considerably distorted and  $S_\alpha$ is rather small. Nevertheless, we
recognize non-negligible cluster formation probability around the surface region of the nucleus 
($a\simeq 4$ fm), which indicates the duality of the shell and cluster as discussed in
Ref. \cite{Chiba2016a}. In terms of the $SU(3)$ shell model,  the $0^+_1$ and $1^-_1$ states 
correspond to the $0\hbar\omega$ and $1\hbar\omega$ configurations, and hence, the nodal quantum
number $n$ of RWAs  should be equal to $n=(N-l)/2$ where $N$ is 8 and 9 for the $0^+_1$ and $1^-$
states, respectively. We clearly see the calculated RWAs follow this relationship. Namely, for
example, the $0^+\otimes 0^+$ RWA has four nodes, while the $2^+\otimes 2^+$ RWA has three.  

Compared to the $0^+_1$ and $1^-_1$ states, the $0^+_6$ and $1^-_{8,9,10}$ states have developed
cluster structure. It is confirmed from their larger $S_\alpha$ and RWAs stretched to the
outward. Different from the $0^+_1$ and $1^-_1$ states, their RWAs do not follow the relationship
of $n=(N-l)/2$. This is due to the mixing with other cluster and non-cluster configurations, which
disturbs the behavior of RWAs. Indeed, we see that RWAs, in particular those of the $1^-_{8,9,10}$
states, show irregular behavior in the inner and outer regions.  
This is consistent with the fact that the ${}^{24}{\rm Mg}+\alpha$ configuration is strongly mixed
with non-cluster configurations and does not appear as a single state but appears as the
$1^-_{8,9,10}$ states in this energy region. It is also noted that the RWAs in the 
$j^\pi\otimes l^{(-)^l}=2^+\otimes l^{(-)^l}$ and $4^+\otimes l^{(-)^l}$ channels are as large as
those of the $0^+\otimes l^{(-)^l}$ channels, which reveals that the rotational excitation of
$^{24}{\rm Mg}$ is coupled to the inter-cluster motion, because of the large deformation of
$^{24}{\rm Mg}$. 

It must be emphasized that the RWAs shown in Fig. \ref{fig:rwa_si28} are hardly obtained by the
ordinary method because large computational cost is demanded. Thus, the Laplace expansion method
realizes the accurate and detailed analysis of the clustering based on the RWAs, which is
indispensable to discuss the clustering in heavier mass nuclei and unstable nuclei.

\section{Summary}\label{sec:summary}
In summary, we presented a new method for the RWA calculation named Laplace expansion method. This
method is based on the Laplace expansion and the analytical separation of the center-of-mass wave
function, and applicable to the Brink-Bloch and AMD wave functions. 
The method enables the calculation of RWA for the different size clusters and deformed clusters
without any approximations. Furthermore, it allows the use of the GCM wave function for the
cluster wave functions, which enables to  calculate the RWA of the non-conventional clusters such
as $^{6}{\rm He}$. Despite of these advantages, the method does not require large computational
cost except for the heavy mass clusters. 

Using the Laplace expansion method, we calculated the RWAs of the $^{16}{\rm O}+\alpha$ clustering
as an example of the unequal size clusters. It was found that the $^{16}{\rm O}+\alpha$ RWA
calculated by using unequal size clusters tends to be smaller than the common size case, and the
difference amounts to 10-20\%. We also presented the RWAs of the $^{24}{\rm Mg}+\alpha$ clustering
as an example of the deformed clusters. It was shown that the RWAs are considerably distorted,
because of the mixing with the cluster and non-cluster configurations. The RWAs also showed that
the rotational excitation of $^{24}{\rm Mg}$ is coupled to the inter-cluster motion, because of the
large deformation of $^{24}{\rm Mg}$. Thus, the Laplace expansion method enables the calculation
of RWA for various cluster systems, and we expect it will be very helpful for the study of the
clustering in heavier mass nuclei and unstale nuclei.  

\section*{Acknowledgment}

Insert the Acknowledgment text here.


\bibliographystyle{ptephy}
\bibliography{RWA}
%

%
%
%
%
%
%

\appendix

\section{Calculataion of the overlaps needed in the Laplace expansion method}\label{app:appa}
Here, we explain the calculation of the overlaps defined by Eqs. \eqref{eq:ov1}, \eqref{eq:ov2}
and \eqref{eq:ov3}.
For arbitrary matrix $\Gamma$, Eq. \eqref{eq:ov1} is calculated by the numerical integration, 
\begin{align}
 \chi_{lm_l}(a;i_1,...,i_A) = \left(\frac{|2\Gamma|}{\pi^3}\right)^{1/4}
 \int d\hat{r}Y^*_{lm_l}(\hat r)\exp\left\{-(\bm r - \bm z)^T\Gamma(\bm r - \bm z)\right\}. 
 \label{eq:ov4}
\end{align}
However, when $\chi(\bm r;i_1,...,i_A)$ is a spherical Gaussian, {\it i.e.}, when the matrix
$\Gamma$ is proportional to the identity matrix $I$ as $\Gamma=\gamma I$, Eq. \eqref{eq:ov4} has 
a simple analytical form,
\begin{align}
 \chi_{lm_l}(r;i_1,...,i_A) = 4\pi i_{l}(2\gamma rz)e^{-\gamma(r^2+z^2)}
\frac{z^lY_{lm_l}^*(\hat z)}{z^l},
\end{align}
where $i_l(2\gamma rz)$ denotes the regular modified spherical Bessel function. The complex
variable $z$ is defined as $z=\sqrt{\bm z\cdot \bm z}$ and $z^lY_{lm_l}^*(\hat z)$ should be
calculated in its Cartesian representation. 

The overlap $ N_{m_1}^{j_1\pi_1}(i_1,...,i_{C_1})$ defined by Eq. \eqref{eq:ov2} is calculated as
follows. For simplicity,  we first assume that the wave function $\Phi^{j_1\pi_1}_{m_1C_1} $  in
the reference state is also represented by a single projected AMD wave function,  
\begin{align}
 \Phi_{C_1}^{AMD} &= \Phi_{C_1}^{cm}\Phi_{C_1}^{int}=\frac{1}{\sqrt{C_1!}} 
 \left| \begin{matrix}
	 \braket{\bm{r}_1|\phi_{i_1}} &  \dots & \braket{\bm{r}_1|\phi_{i_{C_1}}} \\
	 \vdots &  \ddots & \vdots\\
	 \braket{\bm{r}_{C_1}|\phi_{i_1}} & \dots & \braket{\bm{r}_{C_1}|\phi_{i_{C_1}}} 
	\end{matrix}\right|,\label{eq:appwf1}\\
 \Phi_{C_1}^{cm}&= \left(\frac{|2C_1m|}{\pi^3}\right)^{1/4}
 \exp \left\{ -C_1\bm{r}^T m \bm{r} \right\},\\
 \Phi^{j_1\pi_1}_{m_1C_1} &=
 \frac{1}{\sqrt{n^{j_1\pi_1}_{k_1}}}\hat{P}^{j_1}_{m_1k_1}\hat{P}^{\pi_1}
 \Phi_{C_1}^{int},\quad
 n^{j_1\pi_1}_{k_1} = \braket{\Phi_{C_1}^{int}|\hat{P}^{j_1\pi_1}_{k_1k_1}|\Phi_{C_1}^{int}}.
\end{align}
Then, the overlap is given as 
\begin{align}
 N_{m_1}^{j_1\pi_1}(i_1,...,i_{C_1}) = \frac{2j_1+1}{8\pi^2\sqrt{n^{j_1\pi_1}_{k_1}}}
 \int d\Omega D^{j_1*}_{k_1m_1}(\Omega)
 \braket{\Phi_{C_1}^{int}|\hat{P}^{\pi_1}\hat{R}(\Omega)|\Psi_{C_1}^{int}(i_1,...,i_{C_1})},
 \label{eq:ov6}
\end{align}
where the integration over Euler angles are numerically calculated. To calculate the integrand in
Eq. \eqref{eq:ov6}, we introduce an AMD wave function $\widetilde{\Psi}^{AMD}_{C_1}$
\begin{align}
 \widetilde{\Psi}_{C_1}^{AMD}(i_1,...,i_{C_1}) 
 &= \widetilde{\Psi}_{C_1}^{cm}\Psi_{C_1}^{int}(i_1,...,i_{C_1})=\frac{1}{\sqrt{C_1!}} 
 \left| \begin{matrix}
	 \braket{\bm{r}_1|\widetilde{\psi}_{i_1}} &  \dots 
	 & \braket{\bm{r}_1|\widetilde{\psi}_{i_{C_1}}} \\
	 \vdots &  \ddots & \vdots\\
	 \braket{\bm{r}_{C_1}|\widetilde{\psi}_{i_1}} & \dots 
	 & \braket{\bm{r}_{C_1}|\widetilde{\psi}_{i_{C_1}}} 
	\end{matrix}\right|,\label{eq:appwf2}\\
 \braket{\bm{r}|\widetilde{\psi}_i} &= \left(\frac{|2M|}{\pi^3}\right)^{1/4}
 \exp \left\{ -(\bm{r}-\bm{Z}'_i)^T M (\bm{r}-\bm{Z}'_i) \right\} 
 (\alpha_i \chi_\uparrow + \beta_i \chi_\downarrow) \eta_i,\\
 \widetilde{\Psi}_{C_1}^{cm}&= \left(\frac{|2C_1M|}{\pi^3}\right)^{1/4}
 \exp \left\{ -C_1\bm{R}^T_{C_1} M \bm{R}_{C_1} \right\},\\
 \bm Z_i' &= \bm Z_i - \frac{1}{C_1}\sum_{j\in{i_1,...,i_{C_1}}} \bm Z_j, \quad i\in \{i_1,...,i_{C_1}\},
\end{align}
where that the the Gaussian centroids are shifted from the original ones so that the
center-of-mass wave function is located at the origin of the coordinate system. Note that this
shift does not change the internal wave function. Therefore, the internal wave function of
$\widetilde{\Psi}^{AMD}_{C_1}$ is same with the ket state of the integrant in Eq. \eqref{eq:ov6}.  
Using Eqs. \eqref{eq:appwf1} and \eqref{eq:appwf2}, the overlap of AMD wave functions is
calculated as,  
\begin{align}
 \braket{\Phi_{C_1}^{AMD}|\hat{P}^\pi\hat{R}(\Omega)|\Psi_{C_1}^{AMD}}
 = \braket{\Phi_{C_1}^{int}|\hat{P}^\pi\hat{R}(\Omega)|\Psi_{C_1}^{int}}
  \braket{\Phi_{C_1}^{cm}|\hat{P}^\pi\hat{R}(\Omega)|\Psi_{C_1}^{cm}}. \label{eq:ov7}
\end{align}
Here the calculation of the left hand side of Eq. \eqref{eq:ov7} is straightforward, and 
the overlap of the center-of-mass wave functions is analytically calculated as,
\begin{align}
 \braket{\Phi_{C_1}^{cm}|\hat{P}^\pi\hat{R}(\Omega)|\Psi_{C_1}^{cm}} &=
 \left(\frac{|2m||2M|}{|M'+m|^2}\right)^{1/4}, \\
 M' &= R^T(\Omega)MR(\Omega),
\end{align}
where $R(\Omega)$ is a $3\times 3$ rotation matrix which satisfies 
$\hat{R}(\Omega)\bm R =R(\Omega) \bm R$. Therefore, the integrand is proportional to the overlap
of AMD wave functions. 
\begin{align}
 \braket{\Phi_{C_1}^{int}|\hat{P}^\pi\hat{R}(\Omega)|\Psi_{C_1}^{int}}
 =   \left(\frac{|M'+m|^2}{|2m||2M|}\right)^{1/4}
 \braket{\Phi_{C_1}^{AMD}|\hat{P}^\pi\hat{R}(\Omega)|\Psi_{C_1}^{AMD}}. \label{eq:ov8}
\end{align}

It is clear that when $\Phi_{C_1}^{j_1\pi_1}$ and/or $\Psi_{C_1}^{j_1\pi_1}$ are not single AMD
wave function, but GCM wave function, the integrand is a superposition of Eq. \eqref{eq:ov8}. 

\section{An ordinary method for RWA calculation}\label{app:appb}
For the sake of the self-containdness, we explain an ordinary method for RWA calculation
\cite{Horiuchi1972,Horiuchi1977,Kanada-Enyo2003a,Kimura2004a} which is often used in the cluster
models and AMD, and derive Eqs. \eqref{eq:ord1}, \eqref{eq:ord2}, \eqref{eq:ord3} and
\eqref{eq:ord4}.   
We start from a set of the Brink-Bloch type wave functions given in Eq. \eqref{eq:wfbb2}, 
\begin{align}
 \Phi^{BB}_{j_1\pi_1m_1j_2\pi_2m_2}(S_p) &=n_0\mathcal A
 \left\{\hat{P}^{j_1}_{m_1k_1}\hat{P}^{\pi_1} \Phi_{C_1}\left(-\frac{C_2}{A}\bm S_p\right) 
 \hat{P}^{j_2}_{m_2k_2}\hat{P}^{\pi_2} \Phi_{C_2}\left(\frac{C_1}{A}\bm S_p\right)\right\},\\
 n_0&=\sqrt{\frac{C_1!C_2!}{A!}},\quad\bm S_p = (0,0,S_p),
\end{align}
where $\Phi_{C_1}\left(-{C_2}/{A}\bm S_p\right)$ and $\Phi_{C_2}\left({C_1}/{A}\bm S_p\right)$
are the wave functions for clusters with masses $C_1$ and $C_2$, and placed with the
inter-cluster distance $S_p$. They are respectively projected to $j^{\pi_1}$ and
$j^{\pi_2}$. By assuming that $\Phi_{C_1}$ and $\Phi_{C_2}$ are the $SU(3)$ shell model wave 
functions without any particle-hole excitations and have the common oscillator parameter
$\hbar\omega=2\hbar^2\nu/m$, their internal and center-of-mass wave functions can be analytically
separated,  
\begin{align}
 \Phi_{C_1}\left(-\frac{C_2}{A}\bm S_p\right) &= \Phi_{C_1}^{int}\Phi_{C_1}^{cm}, \quad
 \Phi_{C_1}^{cm} = \left(\frac{2C_1\nu}{\pi}\right)^{3/4}
 \exp\left\{-C_1\nu(\bm R_{C_1} + \frac{C_2}{A}\bm S_p)^2\right\}, \\
 \Phi_{C_2}\left(\frac{C_1}{A}\bm S_p\right) &= \Phi_{C_2}^{int}\Phi_{C_2}^{cm},\quad
 \Phi_{C_2}^{cm} = \left(\frac{2C_2\nu}{\pi}\right)^{3/4}
 \exp\left\{-C_2\nu(\bm R_{C_2} - \frac{C_1}{A}\bm S_p)^2\right\}, 
\end{align}
where $\bm R_{C_1}$ and  $\bm R_{C_2}$ are the center-of-mass coordinates of clusters defined by 
Eq. \eqref{eq:rc}.  In a similar way to Eq. \eqref{eq:cmdcmp1}, we rewrite the product of the
center-of-mass wave functions 
as follows,
\begin{align}
 \Phi_{C_1}^{cm} \Phi_{C_2}^{cm} &= \Phi_{A}^{cm}\chi(\bm r), \\
 \Phi_{A}^{cm} &= \left(\frac{2A\nu}{\pi}\right)^{3/4}\exp\left\{-A\nu R^2\right\}, \\
 \chi(\bm r) &= \left(\frac{2\gamma}{\pi}\right)^{3/4}
 \exp\left\{-\gamma(\bm r - \bm S_p)^2\right\}, \quad \gamma = \frac{C_1C_2}{A}\nu
 \label{eq:rel1}. 
\end{align}
Here $\bm R$ and $\bm r$ are the center-of-mass coordinate of $A$-body system and the inter-cluster
coordinate defined by Eqs. \eqref{eq:rc0} and \eqref{eq:rc2}, respectively. Note that the
oscillator parameters of clusters should be the same. Otherwise the decomposition to the
center-of-mass and relative coordinates is not straightforward. Since the relative wave function
Eq. \eqref{eq:rel1} is the coherent state of HO except for a phase factor, it is represented by a
superposition of the HO wave functions \cite{Horiuchi1977},  
\begin{align}
 \chi(\bm r) &= \sum_{Nl} a_{Nl}(S_p) R_{Nl}(r)Y_{l0}(\hat{r}), \\
 a_{Nl}(S_p) &= (-)^{(N-l)/2}\sqrt{\frac{(2l+1)N!}{(N-l)!!(N+l+1)!!}}
 \frac{(\gamma S^2_p)^{N/2}}{\sqrt{N!}}e^{-\gamma S_p^2/2},
\end{align}
where $R_{Nl}(r)$ is the radial wave function of HO and $N$ denotes the principal quantum number.
With these equations, Brink-Bloch type wave function is rewritten as follows,
\begin{align}
 \Phi^{BB}_{j_1\pi_1m_1j_2\pi_2m_2}(S_p) = \Phi_{A}^{cm}\sum_{Nl}a_{Nl}(S_p)n_0\mathcal A
 \Set{R_{Nl}(r)Y_{l0}(\hat r)\Phi^{j_1}_{C_1m_1}\Phi^{j_2}_{C_2m_2}}.
\end{align}
Then, by using the property of the angular momentum projector $P^{J}_{MK}\ket{JK} =\ket{JM}$ and
the coupling of angular momenta, we introduce the wave function, 
\begin{align}
 \Phi^{J\pi}_{j_1\pi_1j_2\pi_2j_{12}l}(S_p)&=\frac{2l+1}{2J+1}
 \sum_{m_{12}m_1m_2}C^{Jm_{12}}_{l0m_{12}}\hat{P}^{J}_{Mm_{12}}C^{j_{12}m_{12}}_{j_1m_1j_2m_2}
 \Phi^{BB}_{j_1\pi_1m_1j_2\pi_2m_2}(S_p) \nonumber \\
 &= \sum_{N}a_{Nl}(S_p)n_0\mathcal A
 \Set{R_{Nl}(r) \left[Y_l(\hat r)\left[\Phi^{j_1\pi_1}_{C_1}\Phi_{C_2}^{j_2\pi_2}\right]_{j_{12}} 
 \right]_{JM}},\label{eq:wfbb3}
\end{align}
in which the angular momenta of clusters are coupled to $j_{12}$, and $j_{12}$ is coupled with the
orbital angular momentum of the relative motion $l$ yielding the total angular momentum $J$. 
$\Phi^{j_1\pi_1}_{C_1}$ and $\Phi^{j_2\pi_2}_{C_2}$ denote the projected internal wave functions, 
$\Phi^{j_1\pi_1}_{C_1}=\hat{P}^{j_1}_{m_1k_1}\Phi_{C_1}^{int}$ and
$\Phi^{j_2\pi_2}_{C_2}=\hat{P}^{j_2}_{m_2k_2}\Phi_{C_2}^{int}$.
When the inter-cluster distance $S_p, \quad p=1,...,p_{max}$ is dense discretized, and maximum
(minimum) distance is chosen to be large (small) enough, a set of wave functions given by
Eq. \eqref{eq:wfbb3} should span the complete set for the $C_1+C_2$ cluster states with
above-mentioned angular momentum coupling, {\it i.e.},
\begin{align}
 &\sum_{pq}\ket{\Phi^{J\pi}_{j_1\pi_1j_2\pi_2j_{12}l}(S_p)}B^{-1}_{pq}
 \bra{\Phi^{J\pi}_{j_1\pi_1j_2\pi_2j_{12}l}(S_q)}\simeq 1, \label{eq:wfbb4}\\
 &B_{pq} = \braket{\Phi^{J\pi}_{j_1\pi_1j_2\pi_2j_{12}l}(S_p)|
 \Phi^{J\pi}_{j_1\pi_1j_2\pi_2j_{12}l}(S_q)}.
\end{align}
Inserting Eq. \eqref{eq:wfbb4} into the definition of the RWA, we get
\begin{align} 
 y^{J\pi}_{j_1{\pi_1}j_2{\pi_2}j_{12}l}(a) =& 
 \sqrt{\frac{A!}{(1+\delta_{C_1C_2})C_1!C_2!}}\nonumber\\
 &\times\sum_{pq}
 \Braket{\frac{\delta(r-a)}{r^2}\left[Y_{l}(\hat{r})
 \left[\Phi^{j_1{\pi_1}}_{C_1} \Phi^{j_2{\pi_2}}_{C_2} \right]_{j_{12}}\right]_{JM}|
 \Phi^{J\pi}_{j_1\pi_1j_2\pi_2j_{12}l}(S_p)}\nonumber\\
 &\times B^{-1}_{pq}\braket{\Phi^{J\pi}_{j_1\pi_1j_2\pi_2j_{12}l}(S_q)|\Psi^{J\pi}_{MA}}.
 \label{eq:ord5}
\end{align}
Using the completeness of the HO wave function $\sum_{N}R_{Nl}(r)R_{Nl}(a)=\delta(r-a)/r^2$ and 
Eq. \eqref{eq:wfbb3}, the braket in the second line reads
\begin{align}
 &\Braket{\frac{\delta(r-a)}{r^2}\left[Y_{l}(\hat{r})
 \left[\Phi^{j_1{\pi_1}}_{C_1} \Phi^{j_2{\pi_2}}_{C_2} \right]_{j_{12}}\right]_{JM}|
 \Phi^{J\pi}_{j_1\pi_1j_2\pi_2j_{12}l}(S_p)} =n_0\sum_{NN'}a_{N'l}(S_p)R_{Nl}(a)\nonumber\\
 &\times \Braket{R_{Nl}(r)\left[Y_{l}(\hat{r})
 \left[\Phi^{j_1{\pi_1}}_{C_1} \Phi^{j_2{\pi_2}}_{C_2} \right]_{j_{12}}\right]_{JM}|\mathcal A
 \Set{R_{N'l}(r) \left[Y_l(\hat r)\left[\Phi^{j_1\pi_1}_{C_1}\Phi_{C_2}^{j_2\pi_2}\right]_{j_{12}} 
 \right]_{JM}}}\nonumber\\
 &=n_0\sum_{N}a_{Nl}(S_p)\mu_{Nl}R_{Nl}(a).\label{eq:ord6}
\end{align}
In the last line, we assumed that $\Phi_{C_1}^{j_1}$ and $\Phi_{C_2}^{j_2}$ are eigenstates of the 
principal quantum number $\hat N$. In this case, the braket in the second line is non-zero only
when $N=N'$, and we denote it $\mu_{N}\delta_{NN'}$. 
From Eqs. \eqref{eq:ord5} and \eqref{eq:ord6},  we get 
\begin{align} 
 y^{J\pi}_{j_1{\pi_1}j_2{\pi_2}j_{12}l}(a) =& 
 \frac{1}{\sqrt{1+\delta_{C_1C_2}}}\sum_N \mu_{Nl}\left(\sum_{pq}a_{Nl}(S_p)
 B^{-1}_{pq}\braket{\Phi^{J\pi}_{j_1\pi_1j_2\pi_2j_{12}l}(S_q)|\Psi^{J\pi}_{MA}}
\right)R_{Nl}(a).
\end{align}
Simplifying this equation, we obtain Eqs. \eqref{eq:ord1}, \eqref{eq:ord2}, \eqref{eq:ord3} and
\eqref{eq:ord4}.
\end{document}